\documentclass{emulateapj}

\slugcomment{Submitted for publication in The Astrophysical Journal}
\shorttitle{Magnetic Interactions in NS Binaries}
\shortauthors{Piro, A. L.}

\newcommand{\be}{\begin{eqnarray}}
\newcommand{\ee}{\end{eqnarray}}
\newcommand{\lp}{\left(}
\newcommand{\rp}{\right)}

\newcommand{\freq}{\lp \frac{\omega}{10^2\ {\rm s^{-1}}}\rp}

\begin{document}


\title{Magnetic Interactions in Coalescing Neutron Star Binaries}

\author{Anthony L. Piro}

\affil{Theoretical Astrophysics, California Institute of Technology, 1200 E California Blvd., M/C 350-17, Pasadena, CA 91125; piro@caltech.edu}


\begin{abstract}
It is expected on both evolutionary and empirical grounds that many merging neutron star (NS) binaries are composed of a highly magnetized NS in orbit with a relatively low magnetic field NS. I study the magnetic interactions of these binaries using the framework of a unipolar inductor model. The e.m.f. generated across the non-magnetic NS as it moves through the magnetosphere sets up a circuit connecting the two stars. The exact features of this circuit depend on the uncertain resistance in the space between the stars $\mathcal{R}_{\rm space}$. Nevertheless, I show that there are interesting observational and/or dynamical effects irrespective of its exact value. When $\mathcal{R}_{\rm space}$ is large, electric dissipation as great as $\sim10^{46}\ {\rm erg\ s^{-1}}$ (for magnetar-strength fields) occurs in the magnetosphere, which would exhibit itself as a hard X-ray precursor in the seconds leading up to merger. With less certainty, there may also be an associated radio transient, but this would be observed well past merger ($\sim\ {\rm hrs}$) because of interstellar dispersion. When $\mathcal{R}_{\rm space}$ is small, electric dissipation largely occurs in the surface layers of the magnetic NS. This can reach  $\sim10^{49}\ {\rm erg\ s^{-1}}$ during the final $\sim1\ {\rm s}$ before merger, similar to the energetics and timescales of short gamma-ray bursts. In addition, for dipole fields greater than $\approx10^{12}\ {\rm G}$ and a small $\mathcal{R}_{\rm space}$, magnetic torques spin up the magnetized NS. This drains angular momentum from the binary and accelerates the inspiral. A faster coalescence results in less orbits occurring before merger, which would impact matched-filtering gravitational-wave searches by ground-based laser interferometers and could create difficulties for studying alternative theories of gravity with compact inspirals.
\end{abstract}



\section{Introduction}

The inspiral and coalescence of a neutron star (NS) with another NS or a black hole is generally considered the most promising target for ground-based gravitational wave detectors. It is anticipated that the interferometers LIGO  \citep{abr92,abb09}, Virgo \citep{car99,ace09}, and KAGRA \citep{kur10}  will reach sufficient sensitivity to observe these events within the next few years once they are upgraded to ``advanced'' sensitivity \citep{aba10}. Since tidal interactions cannot enforce synchronization in a coalescing NS binary \citep{bc92}, the associated waveforms are expected to be relatively simple and can be predicted with high accuracy for the inspiral phase. This makes them ideal for using matched-filtering methods for extracting their signal from the noise \citep{cf94}. In addition, these merger events are of interest for their potential electromagnetic signatures. They are the favored progenitors of short gamma-ray bursts \citep[GRBs,][]{bli84,pac86,pac91,eic89} and have been theorized to produce $\sim1\ {\rm day}$ timescale optical transients from the ejection of neutron-rich material \citep{met08,met10}.

Strong magnetic fields are undoubtedly present in a large fraction of these merging systems. In particular, it is expected that in many binaries one NS has a high magnetic field (\mbox{$\sim10^{12}-10^{14}\ {\rm G}$}) and the other has a lower-strength field ($\sim10^8-10^9\ {\rm G}$). Such a situation is anticipated on theoretical grounds from the binary evolution of two massive main-sequence star \citep{bha91}. The more massive star experiences a supernova first and forms a NS. When the less massive star reaches the end of its main sequence, it becomes a red supergiant before going supernova and producing a NS of its own. This transfers angular momentum and mass to the first NS and decreases its magnetic field \citep{bk74,shi89}. The result is an older NS with a magnetic field similar to a recycled millisecond pulsar in a binary with a younger NS that has a magnetic field more typical of a normal radio pulsar. Such a picture has been confirmed in at least one case for the double pulsar PSR J0737-3039 \citep{bur03,lyn04}, in which the fast and slow spinning NSs have dipole fields of $6.3\times10^9\ {\rm G}$ and $1.2\times10^{12}\ {\rm G}$, respectively. In addition, since $\sim10\%$ of pulsars are born with magnetar strength fields \citep[$\sim10^{14}-10^{15}\ {\rm G}$;][]{kt98}, there is the potential for some binary NSs to have extremely strong fields present (although this may depend on what evolutionary scenarios are needed to produce such high magnetic fields).

The effect of magnetic fields on NS coalescence has been considered in a number previous theoretical studies. Numerical simulations have investigated how the fluid dynamics of the NSs are impacted by magnetic fields, especially post merger \citep[see][and references therein]{and08,liu08,gia09,gia11}. Other work has focused on the inspiral phase to understand the interaction of a non-magnetized NS moving through another NS's magnetosphere \citep{lp96,vie96,hl01}. My work here has a focus similar to these latter studies, but with greater emphasis on the time-dependent binary evolution. The asynchronicity of stars during inspiral provides free energy that can be tapped to drive currents and electrical dissipation. The strength and location of this dissipation is strongly influenced by the (uncertain) resistivity of the space between the merging NSs, $\mathcal{R}_{\rm space}$. Even with this large uncertainty, I show that interesting observational and/or dynamical effects occur no matter what value of $\mathcal{R}_{\rm space}$ is chosen.

In \S \ref{sec:unipolar}, I describe the main features of the so-called ``unipolar inductor'' model applied to binary NSs. In \S \ref{sec:torques}, I investigate the interplay between magnetic torques and the loss of angular momentum due to gravitational wave emission. I provide analytic estimates for the resulting level of synchronization and the electrical power dissipation as a function of orbital frequency. In \S \ref{sec:numerical}, I integrate the time evolution of the coalescing binary numerically to solve for the effects of magnetic fields. I especially focus on how the dissipation in the circuit connecting the two NSs is impacted by $\mathcal{R}_{\rm space}$. In \S \ref{sec:discussion}, I conclude with a summary of my main results, provide a discussion of the potential electromagnetic signatures of these magnetic interactions, and highlight many of the outstanding problems that remain in understanding the impact of magnetic fields on inspiralling NSs.

\section{The Unipolar Inductor Model}
\label{sec:unipolar}

Consider a NS-NS binary merging from gravitational wave emission. One NS has a mass $M_1$ and radius $R_1$ and the other has mass $M_2$ and radius $R_2$. When the binary has an separation $a$, corresponding to a frequency $\omega$, the orbital energy is
\be
	E_{\rm orb} = \frac{GM_1M_2}{a}= (G\omega)^{2/3}\frac{M_1M_2}{M^{1/3}},
\ee
where $M=M_1+M_2$. (For this paper I ignore post-Newtonian corrections since they are not crucial to any of the main results.) At a gravitational wave frequency $f=\omega/\pi=10^2\ {\rm Hz}$, which is near the optimum region for detection by ground-based laser interferometers, the energy is $E_{\rm orb}\approx 3\times10^{52}\ {\rm erg}$. In contrast, the rotational energy of $M_1$ when it has spin $\Omega_1$ is
\be
	E_{\rm rot} = \frac{1}{2}I_1\Omega_1^2,
	\label{eq:erot}
\ee
where $I_1=0.35M_1R_1^2$ is the approximate moment of inertia \citep{lp01}. In the extreme limit of being synchronized, $\Omega_1\approx \omega$ and $E_{\rm rot}\approx 6\times10^{49}\ {\rm erg}$. In other words, if merely $\approx0.2\%$ of the orbital energy is put into a NS's spin, this is enough for the NS to be synchronized with the orbit. Furthermore, typical energies for the prompt emission from short GRBs (once corrected for beaming) are $\sim10^{48}-10^{50}\ {\rm erg}$ \citep{nak07}. So again, if only a small fraction of the orbital energy is tapped (and then emitted on a sufficiently short timescale), it may result in an interesting observable signal. In the following work I argue that magnetic effects may provide the coupling to tap this energy.

For the majority of this paper I use the general framework of a unipolar inductor model to assess the interaction of a non-magnetic NS with a magnetic NS. The main ideas of this model have been presented and explored in a variety of astrophysical settings in the past, from the interaction of Jupiter and Io \citep{gl69}, to ultracompact white dwarf binaries \citep{wu02,dal06}, to short orbital period extrasolar Earths \citep{ll12}, to the interaction of a magnetic NS and black hole \citep{ml11}. Below I discuss this model in the context of merging NSs. The main ideas are summarized schematically in Figures \ref{fig:diagram} and \ref{fig:circuit}, and I will be referring to these throughout the discussion.

\subsection{Skin Depth and Voltage Estimates}
\label{sec:voltage}

 \begin{figure}
\epsscale{1.15}
\plotone{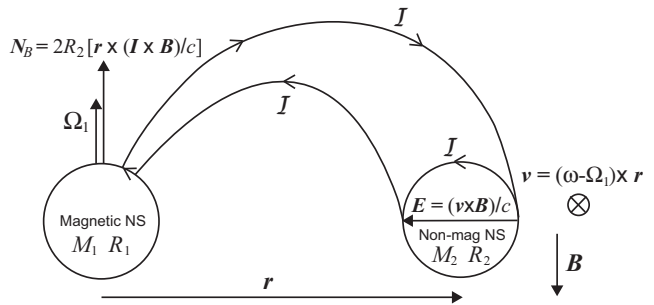}
\caption{Schematic diagram summarizing the main features of the unipolar inductor model applied to binary NSs. The magnetic NS has a dipole field, which threads the non-magnetic NS. The orbital frequency is $\omega$, and the magnetic NS has a spin frequency $\Omega_1$. Therefore, the non-magnetic NS has a relative velocity $\mbox{\boldmath{$v$}}$ with respect to the magnetosphere, which points into the page when $\omega>\Omega_1$. In the frame of the non-magnetic NS, there is an induced electric field $\mbox{\boldmath{$E$}}=(\mbox{\boldmath{$v$}}\mbox{\boldmath{$\times$}}\mbox{\boldmath{$B$}})/c$ (the magnetic field roughly points downward at this location), which drives a current loop $\mathcal{I}$ that is traced out with arrows. For the majority of the loop, the current is forced to flow along the magnetic field lines, with the exception of the surfaces of the two NSs. The interaction of the current over the non-magnetic NS with the dipole field drives a magnetic toque $\mbox{\boldmath{$N_B$}}=2R_2[\mbox{\boldmath{$r$}}\mbox{\boldmath{$\times$}}(\mbox{\boldmath{$\mathcal{I}$}}\mbox{\boldmath{$\times$}}\mbox{\boldmath{$B$}})/c]$. This points in the same direction as $\mbox{\boldmath{$\Omega_1$}}$  when $\omega>\Omega_1$, spinning up the magnetic NS. This diagram only focuses on the upper half of the magnetic interactions. A similar process is repeated in the lower hemisphere which adds to the torque in the same direction.}
\label{fig:diagram}
\epsscale{1.0}
\end{figure}

One NS (labeled as ``1'') has a magnetic moment \mbox{$\mu=BR_1^3$,} where $B$ is its dipole magnetic field strength. I assume that the magnetic moment, the spin axis of each NS, and the spin axis of the orbit are all aligned, as shown in Figure \ref{fig:diagram}. The other NS (labeled as ``2'')  has a negligible magnetic field. The ability of the magnetic field to diffuse into the non-magnetic NS is determined by its conductivity, which is roughly given by \citep{sh53}
\be
	K = \gamma \frac{2(2k_{\rm B}T_e/\pi)^{3/2}}{m_e^{1/2}Ze^2\ln\Lambda},
\ee
where $k_{\rm B}$ is Boltzmann's constant, $T_e$ is the electron temperature, $m_e$ is the electron mass, $e$ is the electron charge, $Z$ is the charge per ion, and $\ln\Lambda$ is the Coulomb logarithm. The factor $\gamma$ depends on the charge per ion $Z$, and can vary between 0.6 (for $Z=1$) and 1 (in the limit $Z\rightarrow\infty$). The characteristic conductivity is therefore $K\approx 1.4\times10^{16}\ {\rm s^{-1}}$, where I take $T_e\approx10^6\ {\rm K}$, $Z=1$, $\gamma\approx0.6$, and $\ln\Lambda\approx10$. 

Since the binary is inspiralling due to gravitational wave emission, there is only a short amount of time for magnetic diffusion to occur.  The rate of angular momentum loss due to gravitational waves is \citep{ll75}
\be
	\dot{J}_{\rm gw} = -\frac{32}{5}\frac{G^3}{c^5}\frac{M_1M_2M}{a^4}J_{\rm orb},
	\label{eq:jgw}
\ee
where $J_{\rm orb} = (Ga/M)^{1/2}M_1M_2$. The inspiral timescale is
\be
	\tau_{\rm gw} &=& \frac{J_{\rm orb}}{3|\dot{J}_{\rm gw}|}= \frac{5}{96} \frac{c^5}{G^{5/3}} \frac{M^{1/3}}{M_1M_2\omega^{8/3}}
	\nonumber
	\\
	&=& 140\freq^{-8/3}\ {\rm s},
	\label{eq:taugw}
\ee
where I have assumed that each NS has a mass of $1.3M_\odot$, and the factor of $1/3$ is because I use the timescale associated with how the orbital frequency is changing ($J_{\rm orb}\propto \omega^{-1/3}$). As described in \citet{jac75}, the skin depth for this timescale is roughly
\be
	\delta \approx \lp \frac{c^2\tau_{\rm gw}}{2\pi K} \rp^{1/2},
	\label{eq:delta1}
\ee
which as a ratio to the non-magnetic NS radius is
\be
	\frac{\delta}{R_2} = 1.2\times10^{-3}K_{16}^{-1/2}\freq^{-4/3},
	\label{eq:delta2}
\ee
where $K_{16}=K/10^{16}\ {\rm s^{-1}}$ and I have used a radius of $R_2=12\ {\rm km}$.

Since this skin depth is small in comparison to the radius, I assume that the non-magnetic NS is roughly a perfect conductor. As described in the Appendix of \citet{hl01}, in this limit the magnetic field must be completely excluded from the non-magnetic NS. This implies that surface currents are induced which produce a magnetic dipole with an opposite orientation with respect to the downward pointing dipole field at the location of the non-magnetic NS. If the non-magnetic NS is also spinning, one must take into account this total magnetic field configuration to correctly understand the interaction. Since the effect of the non-magnetic NS's spin is subdominant to the orbit, and it has already been well-summarized by \citet{hl01}, I ignore it here.

Instead, for this study I focus on the orbital motion of the non-magnetic NS. This induces a surface charge density with a dipole structure pointed toward the magnetic NS
\be
	\Sigma = \mbox{\boldmath{$B$}}\cdot(\mbox{\boldmath{$e$}}_r\mbox{\boldmath{$\times$}}\mbox{\boldmath{$v$}})/4\pi c,
\ee
where $\mbox{\boldmath{$e$}}_r$ is a radial unit vector with respect to the center of the non-magnetic NS, and $\mbox{\boldmath{$v$}}$ is the non-magnetic NS's velocity relative to the magnetosphere. Due to this charge density distribution, there is a net electric field from one side of the non-magnetic NS to the other of $\mbox{\boldmath{$E$}}=(\mbox{\boldmath{$v$}}\mbox{\boldmath{$\times$}}\mbox{\boldmath{$B$}})/c$, and an associated potential difference $\Phi \approx 2R_2|\mbox{\boldmath{$E$}}|$. The e.m.f. when the NSs are at a separation $a$ is therefore
\be
	\Phi \approx \frac{2\mu R_2}{ca^2}(\omega-\Omega_1)
	= \frac{\mu R_2\omega^{7/3}}{c(GM)^{2/3}}\frac{\sigma_1}{\omega},
	\label{eq:phi1}
\ee
where $\sigma_1=2(\omega-\Omega_1)$ is the tidal forcing frequency on the magnetic NS. The numerical value for the voltage is
\be
	\Phi = 3.8\times10^{13}\mu_{31}\lp \frac{\omega}{10^2\ {\rm s^{-1}}} \rp^{7/3}
	\lp\frac{\sigma_1}{\omega} \rp {\rm statvolt},
	\label{eq:phi}
\ee
where $\mu_{31}=\mu/10^{31}\ {\rm G\ cm^3}$. Note that this strength magnetic moment corresponds to a magnetic field of \mbox{$\approx 6\times10^{12}\ {\rm G}$,} which highlights the fact that a surprisingly large potential is possible even for a magnetic field much lower than that typically associated with magnetars.

\subsection{Locations and Rates of Dissipation}

\begin{figure}
\epsscale{1.15}
\plotone{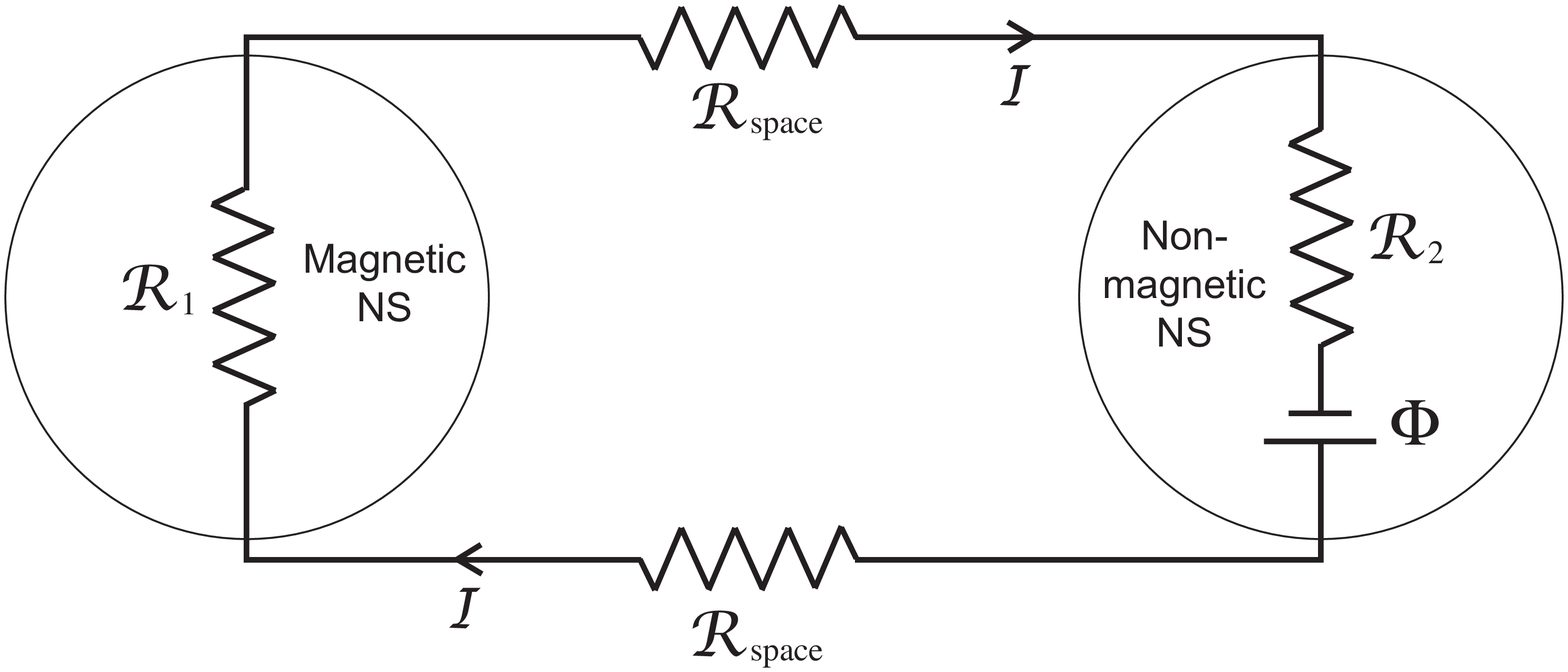}
\caption{Diagram of the circuit formed between the two NSs. The non-magnetic NS has an induced e.m.f. $\Phi$, which drives a current $\mathcal{I}$ over the circuit. Each leg of the circuit has an associated resistivity, which are denoted as $\mathcal{R}_1$, $\mathcal{R}_2$, and $\mathcal{R}_{\rm space}$. The relative size of each of these resistances determine the location and amount of dissipation.}
\label{fig:circuit}
\epsscale{1.0}
\end{figure}

The induced electric field across the non-magnetic NS drives a current over a circuit that connects the two stars, as shown in Figure \ref{fig:circuit}. For a given circuit, the sources of resistance are: the magnetic NS, the non-magnetic NS, and the two regions of space connecting the stars, which are denoted as $\mathcal{R}_1$, $\mathcal{R}_2$, and $\mathcal{R}_{\rm space}$, respectively. Since the resistors are simply connected in series, the current is $\mathcal{I} = \Phi/(\mathcal{R}_1+\mathcal{R}_2+2\mathcal{R}_{\rm space})$. Through each resistor there is an associated rate of dissipation. For example through the magnetic NS,
\be
	\dot{E}_1 = 2\Phi^2\mathcal{R}_1/(\mathcal{R}_1+\mathcal{R}_2+2\mathcal{R}_{\rm space})^2,
\ee
and likewise for each of the other lengths of the circuit (note a factor of two is included because the circuit is repeated in both hemispheres). From this one can see that the relative sizes of the resistances will have a big influence over where the dissipation is greatest and by what amount. For example, if $\mathcal{R}_1\gg\mathcal{R}_2, \mathcal{R}_{\rm space}$, then $\dot{E}_1\approx 2\Phi^2/\mathcal{R}_1$, but if  $\mathcal{R}_{\rm space}\gg\mathcal{R}_1, \mathcal{R}_2$, then $\dot{E}_1\approx \Phi^2\mathcal{R}_1/2\mathcal{R}_{\rm space}^2\ll 2\Phi^2/\mathcal{R}_1$.

For a characteristic vertical length $L$ and area $A$, the resistance is related to the conductivity $K$ by \mbox{$\mathcal{R}=L/AK$.} Now consider the magnetic field lines that simultaneously thread both NSs. For the non-magnetic NS, current flows over its entire surface. But   when the connecting field lines are traced back to the magnetic NS, they only intersect a small fraction of its surface area \mbox{(see Figure \ref{fig:diagram}).} Due to this difference in intersected areas, I approximate that $\mathcal{R}_1\gg \mathcal{R}_2$ and thus ignore $\mathcal{R}_2$ in the rest of my study. Using the geometry of the binary and dipole field, \citet{wu02} estimate
\be
	\mathcal{R}_1 \approx \frac{\mathcal{J}}{R_2K}\lp\frac{H}{\Delta d} \rp \lp\frac{a}{R_1} \rp^{3/2},
\ee
where $\mathcal{J}$ is a geometric factor that depends on the radii of the stars relative to the orbital separation, $H$ is the atmospheric depth at which currents cross magnetic field lines and return back to the non-magnetic NS, and $\Delta d$ is the thickness of the arc-like cross section of the current layer in the magnetic NS's atmosphere. Given the uncertainties in these parameters, I approximate $\mathcal{J}\sim H/\Delta d\sim 1$ for these initial estimates. This results in
\be
	\mathcal{R}_1 = 1.2\times10^{-20}K_{16}^{-1}\lp\frac{\omega}{10^2\ {\rm s^{-1}}} \rp^{-1}{\rm s\ cm^{-1}},
	\label{eq:r1}
\ee
for the resistivity of the magnetic NS.

The other key resistivity is that of the space between the NSs, $\mathcal{R}_{\rm space}$. Given the large potential estimated in equation (\ref{eq:phi}), charged particles are likely accelerated to sufficient energies for pair production. This creation, annihilation, and the subsequent radiation, coupled with other dissipative processes like Alfv\'{e}n wave generation and damping, implies that this space may dominate the resistivity of the circuit \citep[for example, see the discussions of dissipation in magnetar magnetospheres by][]{td95}. In the relativistic limit, one could approximate
\be
	\mathcal{R}_{\rm space}\approx 4\pi/c = 4.2\times10^{-10}\ {\rm s\ cm^{-1}},
\ee
which is over ten orders of magnitude larger than the estimated resistivity of the magnetic NS (eq. [\ref{eq:r1}]). This is the value used by \citet{ml11} for the resistivity of a black hole event horizon \citep[also see][]{dam82,tho86}. But given the complications of what is potentially occurring in the magnetosphere, it is not at all clear that such an extreme resistivity is accurate. So in the interest of understanding the full range of potential solutions, I will explore $\mathcal{R}_{\rm space}$ as a free parameter.

The dissipation can therefore take very different forms depending on the relative resistivities. When the magnetic NS dominates ($\mathcal{R}_1\gg\mathcal{R}_{\rm space}$), most of the dissipation occurs within the surface of the star where the currents run (see \mbox{Figure \ref{fig:diagram}}), mainly producing heat with a rate of
\be
	\dot{E}_1\approx 2.4\times10^{47} \mu_{31}^2 K_{16}
	\lp \frac{\omega}{10^2\ {\rm s^{-1}}} \rp^{17/3}
	\lp\frac{\sigma_1}{\omega} \rp^2{\rm erg\ s^{-1}}.
	\nonumber
	\\
	\label{eq:edotb1}
\ee
This result shows that energy dissipation from magnetic effects is potentially large, but it depends strongly on how asynchronous the magnetic NS is (via the ratio $\sigma_1/\omega$). This is turn is mediated by the competing effects of gravitational wave emission which makes the binary more asynchronous and magnetic torques, which try to synchronize the magnetic NS. I explore this competition in more detail in the next section.

A potentially interesting issue is that as the dissipation heats the surface of the magnetic NS, it would increase its conductivity, which would in turn increase the dissipation, and so on. A more detailed investigation of the surface layer structure is needed to assess what kind of feedback loop could potentially result. Given these uncertainties, I keep $K$ fixed with orbital separation with the caveat that it could in principle change as the binary coalesces.

In the limit when $\mathcal{R}_{\rm space}=4\pi/c\gg\mathcal{R}_1$, the dissipation mainly occurs in the space between the stars with a value
\be
	\dot{E}_{\rm space} \approx 6.8\times10^{36}
	\mu_{31}^2
	\lp \frac{\omega}{10^2\ {\rm s^{-1}}} \rp^{14/3}
	\lp\frac{\sigma_1}{\omega} \rp^2{\rm erg\ s^{-1}}.
	\nonumber
	\\
	\label{eq:edotspace}
\ee
A resistivity this high implies a fairly weak observational signal, at least for typical pulsar-strength fields. If the magnetic moment is increased to $\mu\approx 10^{33}\ {\rm G\ cm^3}$, as is relevant for magnetars, and the binary is close to merger, then this dissipation can be greater than $\sim10^{45}\ {\rm erg\ s^{-1}}$. This is consistent with the prefactor and scalings found in equation (5) of \citet{hl01}, which I explore further in \S \ref{sec:resist}. So even in the limit of the strongest possible resistance, an observational signature of a merger is possible for magnetar strength fields.

Finally, there is an additional source of dissipation that is not associated with the circuit itself, but should be mentioned. Since the non-magnetic NS is effectively a perfect conductor, the magnetic field lines cannot penetrate this star and must be reoriented around it, which requires work. Alternatively, one can also think of this in terms of the current that is generated to produce a magnetic dipole in the opposite orientation to the vertical magnetic field (as was discussed in \S \ref{sec:voltage}). The rate of work done to move the field lines is roughly just the flux of energy density in magnetic field lines that are intercepted by the orbit of the non-magnetic NS,
\be
	\dot{E}_{\rm orb} &\approx& \frac{\mu^2}{8\pi a^6}\times \pi R_2^2 a(\omega-\Omega_1) = \frac{\mu^2R_2^2\omega^{13/3}}{16(GM)^{5/3}}\frac{\sigma_1}{\omega},
	\nonumber
	\\
	&=&2.5\times10^{37}\mu_{31}^2
	\lp \frac{\omega}{10^2\ {\rm s^{-1}}} \rp^{13/3}
	\lp\frac{\sigma_1}{\omega} \rp{\rm erg\ s^{-1}}.
	\nonumber
	\\
\ee
Although this is potentially comparable to the dissipation in the space between the NSs (eq. [\ref{eq:edotspace}]), it is less important because it does not have a corresponding observational signal. It does sap energy from the orbit, but this is a negligible effect in comparison to the gravitational wave emission. Since this will not appreciably alter the inspiral, I ignore it for the remainder of this work.

\section{Magnetic Torques and Dissipation Estimates}
\label{sec:torques}

As shown in Figure \ref{fig:diagram}, the interaction of the current across the non-magnetic NS with the dipole field provides a torque on the magnetic NS, $N_B$, which spins it up when $\omega>\Omega_1$. The total magnetic torque from both hemispheres is
\be
	N_B = \frac{4\mu R_2\mathcal{I}}{ca^2} =\frac{4\Phi^2}{\sigma_1(\mathcal{R}_1+2\mathcal{R}_{\rm space})}.
\ee
In the limit that $\mathcal{R}_1\gg \mathcal{R}_{\rm space}$, $N_B\approx 2\dot{E}_1/\sigma_1$. Similar expressions to this are derived in \citet{wu02} and \citet{dl04}. Note that $N_B\propto \sigma_1\propto \omega-\Omega_1$, so that indeed $N_B>0$ (spinning up the magnetic NS) when  $\omega>\Omega_1$.

The orbital angular momentum of the binary, which is given by $J_{\rm orb} = (Ga/M)^{1/2}M_1M_2$, evolves as
\be
	\dot{J}_{\rm orb} = \dot{J}_{\rm gw} -N_B,
	\label{eq:gw}
\ee
where $\dot{J}_{\rm gw}$ is given in equation (\ref{eq:jgw}). I ignore any tidal effects since they cannot significantly synchronize the binary \citep{bc92}. For solid-body rotation, the spin of the magnetic NS evolves as
\be
	\frac{d}{dt}\lp I_1\Omega_1 \rp \approx N_B +N_p,
	\label{eq:angularmomentum}
\ee
where
\be
	N_p = -\frac{\mu^2\Omega_1^3}{6c^3}.
\ee
is the torque from pulsar dipole spindown.

Before solving the time evolution of the system numerically in \S \ref{sec:numerical}, it is helpful to consider some analytic estimates that highlight the principal effects introduced by the inclusion of  magnetic torques. There are a few key timescales that will be important in determining the evolution of the system. The first is the gravitational wave timescale, which was already presented in \mbox{equation (\ref{eq:taugw}).} Another key timescale is that associated with the magnetic torquing
\be
	\tau_B &=&\frac{ \sigma_1I_1}{N_B}
	= \frac{\omega^2I_1\mathcal{R}_1}{4\Phi^2}\lp\frac{\sigma_1}{\omega} \rp^2\lp 1+\frac{2\mathcal{R}_{\rm space}}{\mathcal{R}_1} \rp
	\nonumber
	\\
	&=& 27\mu_{31}^{-2}K_{16}^{-1}\lp \frac{\omega}{10^{2}\ {\rm s^{-1}}}\rp^{-11/3}
	\lp 1+\frac{2\mathcal{R}_{\rm space}}{\mathcal{R}_1} \rp
	{\rm s}.
	\nonumber
	\\
	\label{eq:tb}
\ee
Finally there is the timescale for spindown from magnetic dipole emission,
\be
	\tau_p &=& \frac{\Omega_1I_1}{|N_p|} = \frac{6c^3I_1}{\mu^2\Omega_1^2}
	\nonumber
	\\
	&=& 6.7\times10^3\mu_{31}^{-2}\lp\frac{\Omega_1}{10^2\ {\rm s^{-1}}} \rp^{-2}{\rm yrs}.
\ee
Even if the magnetic NS is nearly synchronized to the orbit, the dipole spindown is negligible, so I ignore it for the remainder of this study.

The evolution of the system is subject to the competing effects of the gravitational wave emission promoting asynchronicity (thus increasing $\sigma_1$) and the magnetic torques trying to synchronize the magnetic NS (forcing $\sigma_1$ toward zero). This drives the system toward an equilibrium where $d\sigma_1/dt\approx0$ \citep[as was found for the competition between tidal torques and gravitational waves in white dwarf binaries by][]{pir11}. The time derivative of the tidal forcing frequency is
\be
	\frac{d\sigma_1}{dt}=2\lp \frac{d\omega}{dt} - \frac{d\Omega_1}{dt}\rp  = 2\lp \frac{\omega}{\tau_{\rm gw}} -  \frac{\sigma_1}{\tau_B} \rp.
\ee
Setting $d\sigma_1/dt\approx0$, I estimate the steady-state tidal forcing frequency to be
\be
	\frac{\sigma_1}{\omega}  &\approx& \frac{\tau_B}{\tau_{\rm gw}}
	\nonumber
	\\
	&=&0.19 \mu_{31}^{-2}K_{16}^{-1}\lp \frac{\omega}{10^{2}\ {\rm s^{-1}}}\rp^{-1}
	\lp 1+\frac{2\mathcal{R}_{\rm space}}{\mathcal{R}_1} \rp.
	\label{eq:eq}
\ee
For a non-spinning NS, the ratio would be $\sigma_1/\omega = 2$, so this shows that the magnetic torques act to make the magnetic NS more synchronous with the orbit.

This result only holds when $\tau_B\lesssim 2\tau_{\rm gw}$, and therefore magnetic torques are only important if the field exceeds a critical strength of
\be
	B_{\rm crit} = 1.8\times10^{12}K_{16}^{-1/2}
	\lp \frac{\omega}{10^{2}\ {\rm s^{-1}}}\rp^{-1/2}
	\nonumber
	\\
	\times
	\lp 1+\frac{2\mathcal{R}_{\rm space}}{\mathcal{R}_1} \rp^{1/2}
	{\rm G}.
	\label{eq:bcrit}
\ee
This critical field becomes smaller with higher orbital frequencies ($B_{\rm crit}\propto\omega^{-1/2}$), showing that magnetic torques increasingly play a role as the NSs get closer to merger.

Whether or not magnetic torques are important, there will be a large electrical power dissipation on the magnetic NS as long as $\mathcal{R}_{\rm space}$ is negligible.  At early times when \mbox{$B\lesssim B_{\rm crit}$,} I substitute $\sigma_1/\omega\approx2$ into equation (\ref{eq:edotb1}) to find
\be
	\dot{E}_1\approx 9.6\times10^{47} \mu_{31}^2 K_{16}
	\lp \frac{\omega}{10^2\ {\rm s^{-1}}} \rp^{17/3}
	{\rm erg\ s^{-1}},
	\nonumber
	\\
	B\lesssim B_{\rm crit}
	\label{eq:edotb3}
\ee
When $B\gtrsim B_{\rm crit}$, substituting equation (\ref{eq:eq}) into equation (\ref{eq:edotb1}) results in
\be
	\dot{E}_1 \approx8.7\times10^{45}\mu_{31}^{-2} K_{16}^{-1}
	\lp \frac{\omega}{10^2\ {\rm s^{-1}}} \rp^{11/3}
	{\rm erg\ s^{-1}},
	\nonumber
	\\
	 B\gtrsim B_{\rm crit}
	\label{eq:edotb2}
\ee
where I have assumed in both cases that $\mathcal{R}_{\rm space}\ll\mathcal{R}_1$. The dissipation therefore is expected to follow a broken power law, with $\dot{E}_{\rm B}\propto \omega^{17/3}$ at early times and $\dot{E}_{\rm B}\propto \omega^{11/3}$ at late times. The reason for this flattening is that as the magnetic torque becomes more important at a closer orbital separation, the synchronization of the magnetic NS increases and $\sigma_1/\omega$ decreases. This in turn decreases the induced e.m.f.

\section{Numerical Calculations}
\label{sec:numerical}

To explore the time evolution of the coalescing binary in more detail, I now turn to numerical integrations of equations (\ref{eq:gw}) and (\ref{eq:angularmomentum}) using the prescriptions presented in the previous sections. These confirm many of my analytic estimates and provide more details about time-dependent effects. The integrations continue forward in time until the stars reach the tidal disruption separation \citep[][adding Fishbone's 1973 10 percent strong gravity correction]{kop59}
\be
	a_t \approx 2.4 R_2\lp\frac{M}{M_2} \rp^{1/3},
\ee
where I have assumed $M_2\le M_1$. Unless noted otherwise, in all these calculations I use $M_1=M_2=1.3M_\odot$, $R_1=R_2=12\ {\rm km}$, and fix \mbox{$K=10^{16}\ {\rm s^{-1}}$.} This is because my main focus is understanding how varying the magnetic field and resistivity of the space between the NSs impact the evolution.

\subsection{Results when $\mathcal{R}_{\rm space}\ll\mathcal{R}_1$}

 \begin{figure}
\epsscale{1.2}
\plotone{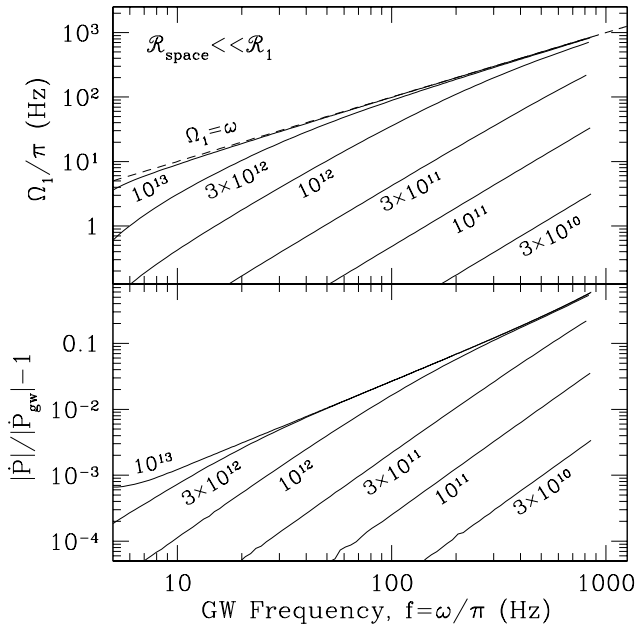}
\caption{Spin of the magnetic NS and the orbital period derivative as a function of the gravitational wave frequency $f=\omega/\pi$ when the resistivity of the space between the NSs is ignored ($\mathcal{R}_{\rm space}\ll\mathcal{R}_1$). The solid lines are labeled by the dipole magnetic field strength (in units of Gauss). For all the calculations I use $M_1=M_2=1.3M_\odot$, $R_1=R_2=12\ {\rm km}$ and fix $K=10^{16}\ {\rm s^{-1}}$. The dashed line marks the condition $\Omega_1=\omega$ in which the magnetized NS is synchronized with the orbit. Magnetic fields stronger than those plotted here simply push the magnetic NS even closer to synchronization.}
\label{fig:spinevolution}
\epsscale{1.0}
\end{figure}

I begin by exploring the case when $\mathcal{R}_{\rm space}\ll\mathcal{R}_1$, so that the resistivity of the space between the stars is effectively negligible. This makes a direction connection with the analytic results from the end of \S \ref{sec:torques}. In the top panel of Figure \ref{fig:spinevolution}, I plot the spin frequency of the magnetized NS as a function of the gravitational wave frequency $f=\omega/\pi$. The solid lines are labeled by the dipole field strength, which varies from $3\times10^{10}\ {\rm G}$ to $10^{13}\ {\rm G}$, from bottom to top. The dashed line is the relation $\Omega_1=\omega$, showing that as the field increases the magnetic torques drive the magnetic NS toward being synchronized. For a field of $10^{12}\ {\rm G}$, synchronization starts becoming important around a frequency of \mbox{$f\approx 200\ {\rm Hz}$.} This is roughly consistent with $B_{\rm crit}$ estimated from plugging $\omega=\pi f\approx 600\ {\rm s^{-1}}$ into \mbox{equation (\ref{eq:bcrit}).}

As the magnetized NS spins up, it could in principle experience rotational instabilities which would result in a triaxial shape and produce additional gravitational waves. These instabilities scale with the rotation parameter $\beta=E_{\rm rot}/|W|$, where $E_{\rm rot}$ is given in equation (\ref{eq:erot}) and I use the prescription given in \citet{lp01} for
\be
	|W| \approx 0.6M_1c^2\frac{GM_1/R_1c^2}{1-0.5(GM_1/R_1c^2)}.
\ee
Dynamical bar-mode instabilities occur for $\beta>0.27$ \citep{cha69}, and secular instabilities for $\beta>0.14$, driven by gravitational radiation reaction or viscosity \citep{ls95}. Even for the latter case, it requires $f\gtrsim 2.8\times10^3\ {\rm Hz}$ for instability, so in practice the binary always merges before sufficiently high spin frequencies can be reached. 

In the bottom panel of Figure \ref{fig:spinevolution}, I plot the fractional difference of the orbital period derivative
\be
	|\dot{P}| = 6\pi \lp\frac{J_{\rm orb}}{GM}\rp^2 \lp \frac{M}{M_1M_2}\rp^3|\dot{J}_{\rm orb}|,
\ee
in comparison to that given from just gravitational wave emission $|\dot{P}_{\rm gw}|$ (which is found by just substituting $|\dot{J}_{\rm gw}|$ for $|\dot{J}_{\rm orb}|$ in the previous expression). In general, because magnetic torques remove angular momentum from the orbit and put it into the magnetized NS, the absolute value of the orbital period derivative increases (i.e., becomes more negative). The maximum effect occurs when the magnetized NS is synchronized with the orbit. The bottom panel of Figure \ref{fig:spinevolution} shows that the period derivative is larger by as much as $\approx10-40\%$ at late times. It should be emphasized that this is the maximum possible deviation. For magnetic fields $\gtrsim10^{13}\ {\rm G}$, the period derivative cannot change by any more because the magnetic NS is already nearly synchronized with the orbit at large separation. The larger period derivative means that the binary experiences less orbits while its inspiral is at frequencies detectable by ground-based interferometers. This will create difficulties in using matched-filtering techniques to pull these binary inspiral signals out of the noise and making reliable parameter estimations \citep{cf94}.

 \begin{figure}
\epsscale{1.2}
\plotone{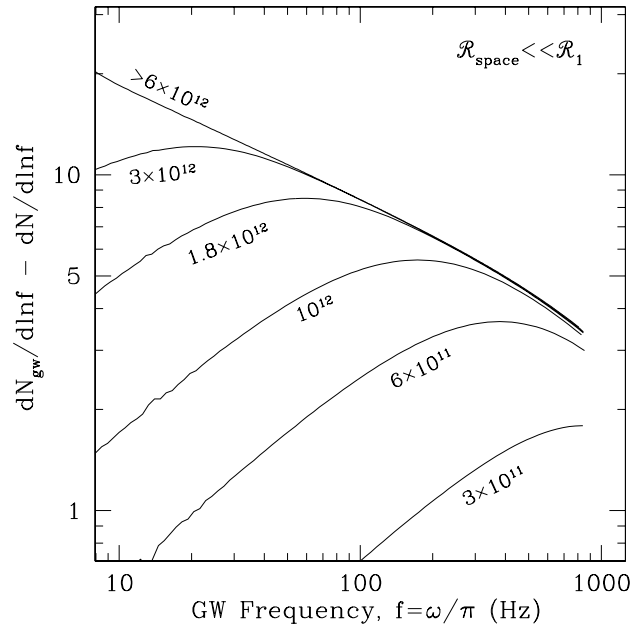}
\caption{The difference in the number of orbits per logarithm orbital frequency due to magnetic torques when $\mathcal{R}_{\rm space}\ll\mathcal{R}_1$. Each solid line is labeled by the magnetic field strength (in units of Gauss). All other quantities are fixed the same as in Figure \ref{fig:spinevolution}. From this presentation, one can see at what frequencies the most number of orbits are being lost. For a magnetic field $>6\times10^{12}\ {\rm G}$, no additional orbits are lost because the magnetic NS is already effectively tidally locked. }
\label{fig:norb}
\epsscale{1.0}
\end{figure}
Although the period derivative changes by the largest amount at high frequencies, there are many more orbits at low frequencies. This means that the largest difference in the number of orbits (which is what is important for matched filtering) may be at earlier times. To better quantify how many fewer orbits are expected and at what times, I plot the differential number of orbits $N$ per logarithm frequency in Figure \ref{fig:norb}, where
\be
	\frac{dN}{d\ln f} = |\dot{P} |^{-1},
\ee
and similarly $dN_{\rm gw}/d\ln f = |\dot{P}_{\rm gw} |^{-1}$. Such a plot shows at what orbital frequencies the most number of orbits are being lost. From Figure \ref{fig:norb}, one can see that as the magnetic field increases, the majority of the orbits are lost at earlier times during the inspiral. For example, at a field strength of $10^{12}\ {\rm G}$, a few orbits are lost around a frequency of $f\approx200\ {\rm Hz}$. Such differences in the inspiral will have to be accounted for in a matched-filtering analysis (which would of course also include post-Newtonian effects that I have omitted for simplicity).

\begin{figure}
\epsscale{1.2}
\plotone{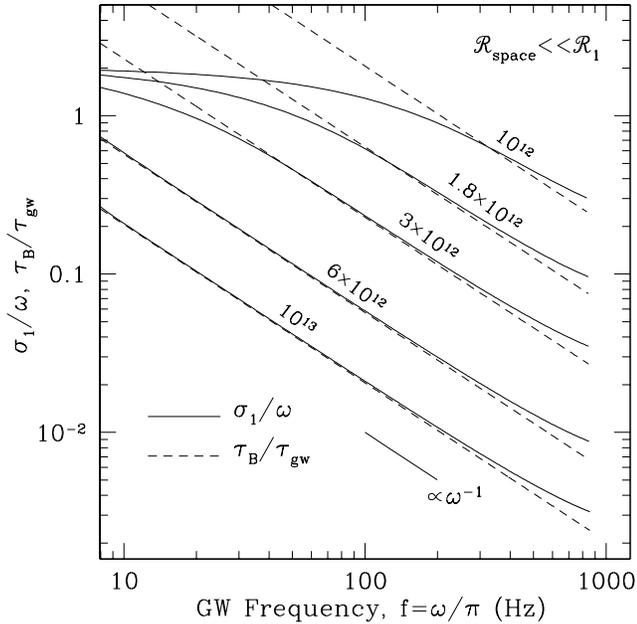}
\caption{The ratio of the tidal forcing frequency to orbital frequency, $\sigma_1/\omega$, as a function of the gravitational wave frequency $f=\omega/\pi$ (solid lines). Each curve is labeled by the magnetic field strength (in units of Gauss), with all other quantities kept fixed with the same values as in Figure \ref{fig:spinevolution}. The dashed curves show the ratio of the magnetic torquing timescale to the gravitational wave times, $\tau_{\rm B}/\tau_{\rm gw}$, for each corresponding magnetic field, which is predicted to be a good estimate for the ratio $\sigma_1/\omega$ and be proportional to $\omega^{-1}$ in \mbox{equation (\ref{eq:eq}).} }
\label{fig:sigma}
\epsscale{1.0}
\end{figure}

I next check the relation estimated in equation (\ref{eq:eq}) for the tidal forcing frequency in Figure \ref{fig:sigma}. In each case at early times $\sigma_1/\omega\approx 2$, which corresponds to a low spin frequency for the magnetized NS. But as the magnetic torque increases at high orbital frequency, $\sigma_1/\omega$ roughly asymptotes to the ratio of the magnetic torquing timescale to the gravitational wave timescale, $\tau_B/\tau_{\rm gw}$, which is plotted as a dashed line. At magnetic field strength $\lesssim10^{12}\ {\rm G}$ the magnetic torques are less important and equation (\ref{eq:eq}) no longer applies.

 \begin{figure}
\epsscale{1.2}
\plotone{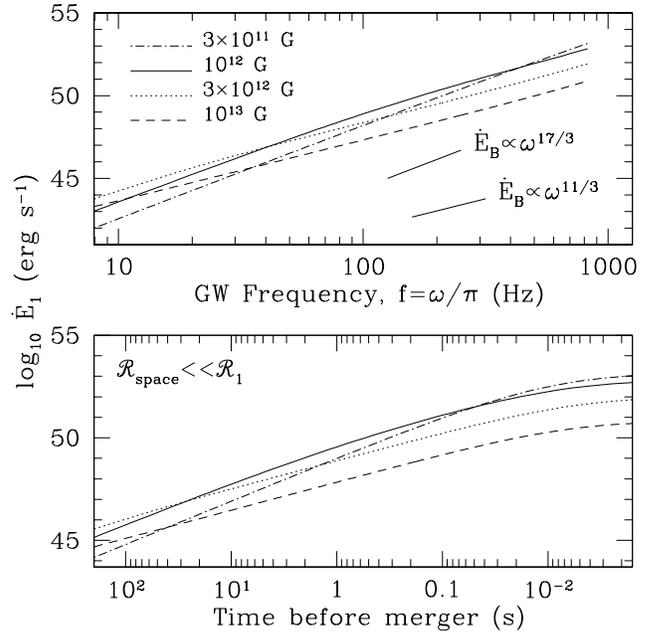}
\caption{The electrical dissipation power plotted as a function of both the gravitational wave frequency $f=\omega/\pi$ (top panel) and the time until merger ($a\approx a_t$, bottom panel). The different line types correspond to different magnetic fields, as indicated by the key in the upper panel.}
\label{fig:power}
\epsscale{1.0}
\end{figure}
In Figure \ref{fig:power}, I calculate the electrical power dissipation on the surface of the magnetic NS, $\dot{E}_1$, for a range of magnetic field strengths. In the top panel I plot this as a function of $f$. This shows a dependence on the frequency that roughly matches the scalings given in equations (\ref{eq:edotb3}) and (\ref{eq:edotb2}) at low and high fields, respectively. Due to the effect of magnetic torques, the highly magnetic NSs show more dissipation at low frequencies and low magnetic field NSs show more dissipation at high frequencies. Although the power can get rather large at $\dot{E}_B\sim 10^{49}-10^{52}\ {\rm erg\ s^{-1}}$, this only last for a short time. In the bottom panel of Figure \ref{fig:power}, I plot these same energy dissipation rates as a function of time until merger ($a\approx a_t$). This shows that a rate of $\gtrsim10^{49}\ {\rm erg\ s^{-1}}$ is only possible for the last $\sim1\ {\rm s}$. Such energetics and timescales are similar to short GRBs, a fact I discuss in more detail in the conclusion.

\subsection{Results when $\mathcal{R}_{\rm space}$ is Non-negligible}
\label{sec:resist}

 \begin{figure}
\epsscale{1.2}
\plotone{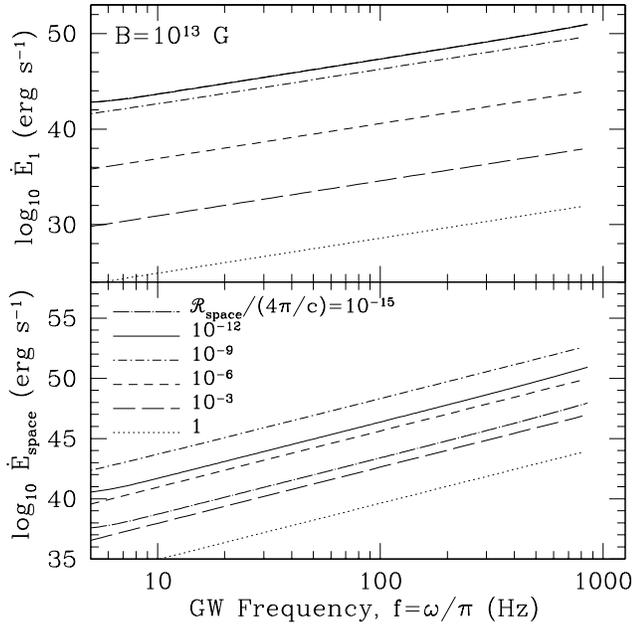}
\caption{The dissipation rates on the surface of the magnetic NS ($\dot{E}_1$, top panel) and in the space between the NSs ($\dot{E}_{\rm space}$, bottom panel) as $\mathcal{R}_{\rm space}$ changes. The value of $\mathcal{R}_{\rm space}$ are in units of $4\pi/c$ as denoted by the key in the bottom panel. For all the calculations I fix $B=10^{13}\ {\rm G}$, and the masses and radii are the same as in Figure \ref{fig:spinevolution}.}
\label{fig:resistivity}
\epsscale{1.0}
\end{figure}
The largest uncertainty in these calculations is the resistivity in the space between the stars $\mathcal{R}_{\rm space}$. Thus far I have been assuming that $\mathcal{R}_{\rm space}$ is negligible, but interesting phenomena potentially occur no matter what value it takes. In Figure \ref{fig:resistivity}, I fix the magnetic field at $10^{13}\ {\rm G}$ and explore the effect of increasing $\mathcal{R}_{\rm space}$ on the dissipation from the magnetic NS $\dot{E}_1$ (top panel), and the dissipation in the space between the stars $\dot{E}_{\rm space}$ (bottom panel). Each line in Figure \ref{fig:resistivity} corresponding to a different value of $\mathcal{R}_{\rm space}$ in units of the maximum resistivity $4\pi/c$ according to the key in the bottom panel. This shows that there is some give and take between these two rates of dissipation, and independent of the exact value, it appears that the dissipation somewhere should be observationally important.

 \begin{figure}
\epsscale{1.2}
\plotone{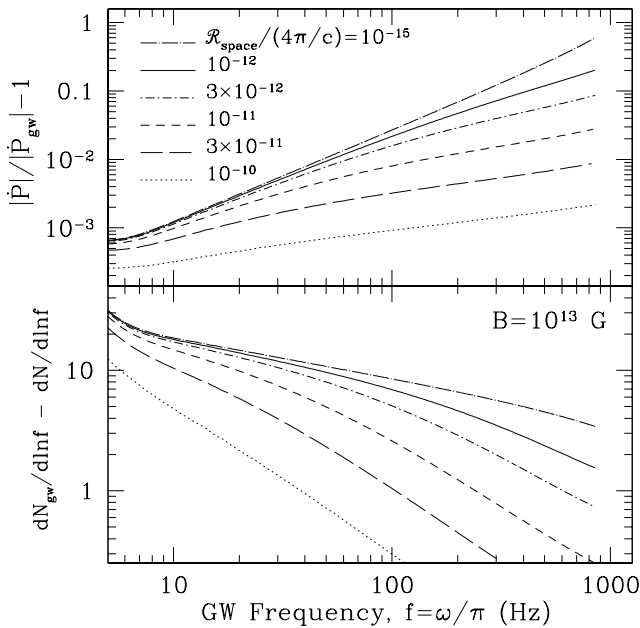}
\caption{The period derivative (top panel) and number of orbits lost (bottom panel), for different values of $\mathcal{R}_{\rm space}$ in units of $4\pi/c$, as labeled. This demonstrates that the differences in the period derivative and number of orbits are strongly dependent on $\mathcal{R}_{\rm space}$. For all the calculations I fix $B=10^{13}\ {\rm G}$, and the masses and radii are the same as in Figure \ref{fig:spinevolution}.}
\label{fig:pdot_resist}
\epsscale{1.0}
\end{figure}

On the other hand, the deviation of the coalescence period derivative from just a point-particle inspiral is strongly effected by $\mathcal{R}_{\rm space}$, as is show in Figure \ref{fig:pdot_resist}. The top panel shows the fractional change of the period derivative. When $\mathcal{R}_{\rm space}$ is small, I effectively get the same large deviation of the period derivative as was found in Figure \ref{fig:spinevolution} before. These differences persists as long as $\mathcal{R}_{\rm space}\lesssim\mathcal{R}_1$. But once these are about equal when $\mathcal{R}_{\rm space}\lesssim 10^{-10}\times4\pi/c$, the period derivative is expected to be consistent with point-particle inspiral. In the bottom panel of Figure \ref{fig:pdot_resist}, I better quantify this in terms of the number of orbits that are lost at a given frequency (similar to what was done before in Figure \ref{fig:norb}).

 \begin{figure}
\epsscale{1.2}
\plotone{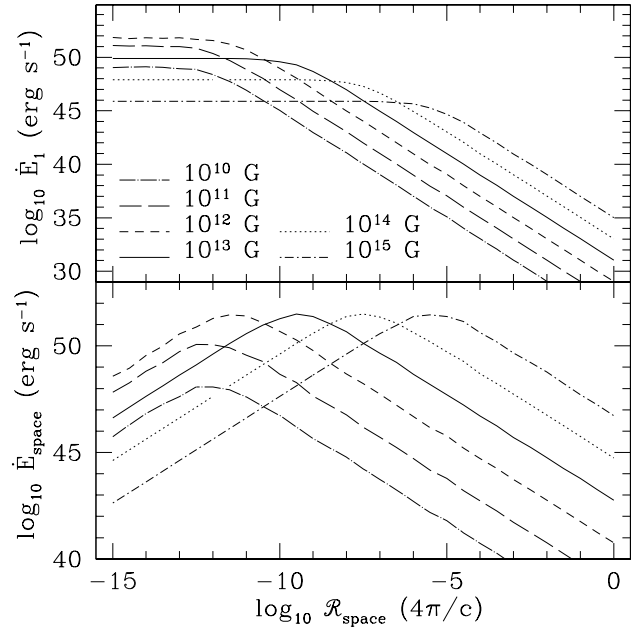}
\caption{The dissipation rates at the moment $\tau_{\rm gw}=0.1\ {\rm s}$ which corresponds to $f\approx480\ {\rm Hz}$, from both the surface of the magnetic NS ($\dot{E}_1$, top panel) and in the space between the NSs ($\dot{E}_{\rm space}$, bottom panel) as $\mathcal{R}_{\rm space}$ changes. For all the calculations the masses and radii are the same as in Figure \ref{fig:spinevolution}.}
\label{fig:lastmoment}
\epsscale{1.0}
\end{figure}

In all cases the peak dissipation rates are reached near the end of the NS inspiral, and this provides the most exciting opportunity for a transient signature coincident with gravitational wave emission. To better quantify what is expect, I plot the dissipation rates at the moment $\tau_{\rm gw}=0.1\ {\rm s}$ in Figure \ref{fig:lastmoment}. This corresponds to an orbital frequency of $\omega=1.5\times10^3\ {\rm s^{-1}}$ or a gravitational wave frequency of $f\approx480\ {\rm Hz}$. This figure again displays the give and take between the two dissipation locations as $\mathcal{R}_{\rm space}$ changes. It also highlights that the most potentially optimistic case comes when $\mathcal{R}_{\rm space}\approx\mathcal{R}_1$. Also note that in the limit $\mathcal{R}_{\rm space}=4\pi/c$ and $B=10^{15}\ {\rm G}$, the dissipation in the space between the NSs $\dot{E}_{\rm space}$ is similar to the estimates in \citet{hl01}.

\section{Discussion and Conclusions}
\label{sec:discussion}

I investigated the inspiral of NS binaries when one NS has a magnetic field. Using a unipolar inductor model, I assessed what typed of currents and associated electrical dissipation rates are expected. The main uncertainty in this analysis was the resistivity between the stars, $\mathcal{R}_{\rm space}$, but for whatever value it takes in nature there are interesting consequences to explore. When $\mathcal{R}_{\rm space}$ is large, electrical dissipation comes predominantly from the space between the NSs. This implies specific observational predictions, which I discuss below. When $\mathcal{R}_{\rm space}$ is small, the competing effects of magnetic torques spinning up the magnetized NS and gravitational wave emission causing the binary to coalesce result in a steady-state asynchronicity. I explored this both analytically and with numerical integrations of the binary time evolution. The loss of angular momentum from the binary that is used to spin up the magnetic NS impacts the dynamics of the inspiral. There is also a large associated dissipation of currents in the surface  of the magnetic NS that may again have observational consequences.

\subsection{Observational Signatures}

When $\mathcal{R}_{\rm space}$ is non-negligible, the dissipation mainly occurs in the space between the stars. This will potentially result in a couple of different signatures, which are analogous to the effects \citet{hl01} describe in some detail, but I quickly summarize the main results here for completeness. The electric field generated around the non-magnetic NS accelerates particles to sufficient energies for pair production, analogous to radio pulsars. In pulsars, the energy extraction is limited to the polar caps, but for the non-magnetic NS, all field lines extend away, so that the polar cap effectively encompasses the whole star. From this \citet{hl01} estimate that for a magnetar a radio flux of $\sim{\rm few}\ {\rm mJy}$ is possible. On the other hand, there are many uncertainties in whether the radio emission will suppressed by the strong magnetic field \citep{um96,aro98} and the shroud of plasma within the magnetosphere \citep{gj69}. Also, it is uncertain what kind of delay interstellar dispersion may create \citep{pal93,lip97}, causing the radio to follow the actual merger by hours or more. Potentially more promising in the case of a large $\mathcal{R}_{\rm space}$ is the energy dissipation of accelerated particles and Alfv\'{e}n waves in the space between the stars. This would create a hard X-ray precursor from a relativistically expanding wind of pairs and photons that would occur $\sim\ {\rm seconds}$ before merger.

In the limit when $\mathcal{R}_{\rm space}$ is small, the majority of the dissipation occurs in currents in the surface of the magnetic NS. This can reach rates of \mbox{$\gtrsim10^{49}\ {\rm erg\ s^{-1}}$} during the final $\sim1\ {\rm s}$ before merger, as shown in Figure \ref{fig:lastmoment}. Although the energetics and timescale of the dissipation are similar to short GRBs, it is not clear what the observational outcome is here. The majority of the dissipation will take place where the field lines connecting to the two NSs intersect the magnetized NS. This presumably is released as heat. If a small fraction of mass is ablated from this energy injection, it could produce a pair fireball that may be like a GRB. The main difficulty with making this connection is that it is not clear what collimates the flow sufficiently to match the observed jets \citep[although there is some evidence that at least some short GRBs have relatively wide jets,][]{gru06}, but this clearly merits a more detailed study.

\subsection{Future Work}

The effect of magnetic torques on binary NS inspiral dynamics deserves further study, especially since matched-filtering is anticipated to play an important role in detecting these events with ground-based interferometers. When the inspiral is altered by magnetic torques, it will likely lead to misestimations of binary parameters. Future work should quantify which parameters are most impacted and what are the sizes of the induced uncertainties. In addition, compact binary coalescences have historically been identified as promising sites for studying the strong-field dynamics of general relativity, and potentially testing for other theories of gravity \citep[see][and references therein]{li12}. Magnetic fields will make such efforts more difficult.

On the other hand, given some of the electromagnetic predictions from these magnetic interactions, there may be correlations between different inspiral signals and X-ray precursors ($\sim{\rm sec}$ before merger) and/or later radio emission ({$\sim{\rm hrs}$ after). This adds to a host of other potential electromagnetic phenomena that have been theorized to be associated with compacter mergers, including crust cracking in the seconds before merger \citep{tsa12}, kilonovae in the day following merger \citep{met08,met10}, and radio transients a few weeks later \citep{np11}. Each of these emission mechanisms probe different, complementary aspects of the merging binary, and when taken together could provide a detailed picture of exactly what type of objects are being detected with gravitational waves.

A potential complication is that throughout this study I have assumed that the less magnetic NS has a completely negligible field. This is a good approximation at late times and for very strong magnetic fields for the magnetic NS. But because a dipole field drops off so rapidly, it does not take a very large separation for each NS to have its own magnetosphere. For example, in the case of the binary pulsar PSR J0737-3039, only for $a\lesssim 6R_1$ will the higher magnetic field really dominate. Prior to this time, a different analysis might be required that takes into account both fields \citep[e.g.,][]{dem04,kas04,lyu04,aro05}.

Finally, another important problem is the coupling between the magnetic field, the magnetic NS's crust, and its core. In the present work I assume that the magnetic torque simply acts over the entire magnetic NS, and that it remains rigidly rotating. But since the field is primarily locked into the crust and the torques are concentrated at the magnetic footpoints (as shown in Figure \ref{fig:diagram}), it is not clear if my assumptions hold in detail. For example, a solid NS crust has an associated shear modulus $\mu_{\rm cr}\sim10^{30}\ {\rm erg\ cm^{-3}}$ \citep[substituting typical values for the density and composition into the calculations by][]{str91}. In comparison, typical torques that we consider have values of $N_{\rm B}\approx 2\dot{E}_B/\sigma_1\sim 10^{46}-10^{50}\ {\rm erg}$. Even when averaged over the entire NS crust, with a typical thickness of $\sim0.1R$, this gives a torque density of $\sim 10^{29}-10^{33}\ {\rm erg\ cm^{-3}}$. Given that the torque is actually exerted on smaller regions of the crust, it seems likely that they may exceed the crust breaking strain \citep{hk09}. A more detailed study of the crustal coupling is therefore needed. This will help develop a better understanding of how the torques are transmitted through the NS. If crust cracking occurs, it may lead to interesting precursor phenomena, such as in the observations reported by \citep{tro10}, and suggested to be due to crustal breaking (albeit by a different mechanism) by \citet{tsa12}.
\\

I thank  Douglas N.C. Lin for discussions of the unipolar inductor model (in a different context), which provided the impetus for this research. I also thank Andrei Beloborodov, Lars Bildsten, Philip Chang, Peter Goldreich, Maxim Lyutikov, Christian Ott, Anatoly Spitkovsky, and David Tsang for helpful comments and feedback on previous drafts. This work was supported through NSF grants AST-0855535 and PHY-1069991, and by the Sherman Fairchild Foundation.


\end{document}